\documentclass[%  
reprint,
groupedaddress,
 amsmath,amssymb,
 aps,
showkeys
]{revtex4-2}
\usepackage{siunitx}
\usepackage{filecontents}
\usepackage{graphicx}% Include figure files
\usepackage{dcolumn}% Align table columns on decimal point
\usepackage{bm}% bold math
\usepackage{soul}
\usepackage{tikz}
\usepackage{pgfplots}
\usepackage{pgfplotstable}
\pgfplotsset{compat=1.18}

\usepackage[svgnames]{xcolor}
\newcommand{\rev}[1]{\textcolor{black}{#1}} % blue = revised text
\begin{document}

\preprint{APS/123-QED}

\title{Gravitational Wave Effects on Radio Spectral Lines of Atomic Hydrogen:\\Hyperfine Splitting and Broadening Mechanisms}

\author{Nontapat Wanwieng}
\email{nontapat@narit.or.th}
\affiliation{National Astronomical Research Institute of Thailand (NARIT), Chiang Mai, 50180,Thailand}

\author{Nithiwadee Thaicharoen}
\affiliation{%
Department of Physics and Materials Science, Faculty of Science,
Chiang Mai University, Chiang Mai 50200, Thailand
}%
\author{Narupon Chattrapiban}
\affiliation{%
Department of Physics and Materials Science, Faculty of Science,
Chiang Mai University, Chiang Mai 50200, Thailand
}%
\author{Apimook Watcharangkool}%
 \email{apimook@narit.or.th}
 \affiliation{National Astronomical Research Institute of Thailand (NARIT), Chiang Mai, 50180,Thailand}
\affiliation{
 Department of Physics and Materials Science, Faculty of Science,
Chiang Mai University, Chiang Mai 50200, Thailand
}%


\date{\today}% It is always \today, today,

\begin{abstract}
We explore the effects of gravitational waves (GWs) on hydrogen's radio spectral lines, focusing on the ground-state hyperfine transition and radiative transitions in highly excited Rydberg states. To analyze GW impacts on hyperfine structure, we derive Maxwell's equations in a gravitational-wave background using linearized gravity and the $3+1$ formalism. Our findings reveal that GWs induce energy shifts in hyperfine magnetic substates, modifying the 21 cm line. However, these energy shifts fall well below the detection limits of current radio astronomical instruments.
For transitions in highly excited states, which produce radio recombination lines (RRL), the influence of GW manifests itself as spectral broadening, with the fractional linewidth for $\mathrm{H}n\alpha$ scaling as $\Delta\nu/\nu_0 \sim n^7\omega^2_{\mathrm{gw}}h(t)$. This suggests that RRLs could serve as probes for ultra-high-frequency GWs, particularly given that Rydberg atoms in the interstellar medium can reach quantum numbers above $n=100$. As an example of possibly detectable high frequency GW source, We investigate GWs emitted during the inspiral of planetary-mass primordial black hole binaries, where GW-induced broadening in RRLs could exceed natural broadening effects. Additionally, we examine the influence of the recently detected stochastic gravitational-wave background on hydrogen spectral lines.
\end{abstract}

 %Use showkeys class option if keyword
\maketitle

\section{Introduction}\label{sec:intro}
Gravitational waves (GWs), as predicted by Einstein’s General Relativity \cite{einstein1916,einstein1918}, are ripples in spacetime that travel at the speed of light. Since their first direct detection by LIGO in 2015 \cite{abbott2016}, GWs have provided unprecedented insights into astrophysical phenomena that were previously inaccessible through traditional electromagnetic observations. More recently, the detection of a stochastic gravitational-wave background (SGWB) by Pulsar Timing Arrays (PTAs) \cite{agazie2023} has expanded our ability to probe the universe on cosmological scales.

GWs cover a broad range of frequencies, each linked to specific astrophysical sources and detection techniques. Nanohertz-frequency GWs are monitored by PTAs, which primarily detect signals from supermassive black hole binaries. The space-based LISA mission will observe the millihertz band, detecting compact binary systems such as white dwarf mergers and extreme-mass-ratio inspirals. Ground-based detectors like LIGO, Virgo, and KAGRA operate in the tens of hertz to several kilohertz range, capturing mergers of stellar-mass black holes and neutron stars. Meanwhile, GWs above 10 kHz remain largely uncharted but could provide insights into exotic sources such as primordial black holes or physics beyond the Standard Model \cite{aggarwal2025}.

Since the current detection technologies rely on laser interferometry, the detectable frequency range is limited by the quantum nature of light (e.g. shot noise). To explore the frequency outside of the range 10 Hz - 10 kHz, a new detection method needs to be employed. The Terrestrial very-long-baseline atom interferometer (TVLBAI) is one such example \cite{TVLBAI}. This matter-wave interferometer is expected to be sensitive to GWs in deci-Hertz band, providing data in a frequency range unreachable by the current detection technology.

For high frequency GWs above $10$ kHz, a detection possibility using cold atoms has been discussed \cite{wanwieng2023}. Note that the effect of the gravity on the spectra of atoms and molecules has been previously investigated by several researchers \cite{parker1980,parker1982,leen1983,siparov2004,Bringmann2023,Kahn2024}, but previously believed to be impossible for a ground-based detection, due to the weakness of the signal. To improve the sensitivity, one may consider atoms in highly excited states, i.e. Rydberg states \cite{Kanno2025}, or by looking at spectrum of atoms near GW sources e.g. interstellar medium.

In low-density regions of the interstellar medium (ISM), ions can capture free electrons at very high quantum numbers (typically $n > 100$) forming Rydberg atoms \cite{gnedin2009}. As these atoms undergo a cascade through lower high-lying states, each transition produces radio recombination lines (RRLs). These lines serve as valuable tracers of ionized environments, including HII regions and the diffuse ISM. 

Advancements in radio astronomy, particularly with high-resolution instruments such as the Square Kilometre Array (SKA) and the Very Large Array (VLA), have significantly enhanced the ability to measure RRLs with greater precision \cite{roelfsema1992,peters2011}. Observations have detected carbon RRLs from states with quantum numbers as high as $n=1500$ in the cool, tenuous medium \cite{stepkin2007}. At such high excited states, these Rydberg atoms become extremely large, sometimes reaching macroscopic sizes, and are consequently fragile. The transitions between high-lying energy levels are highly sensitive to their surroundings, making them useful probes of local conditions such as temperature, density, and magnetic fields. A compelling question is whether these transitions also exhibit sensitivity to gravitational fields, e.g. near black holes or gravitational-wave sources. Investigating this possibility could provide insights into the observable effects of local gravity on atomic transitions in astrophysical environments.

This paper is organized as follows: In Sec. II, we derive the hyperfine structure modifications in the presence of GWs, using electrodynamics in curved spacetime. In Sec. III, we analyze the effects of GWs on hydrogen Rydberg states, including quantum mechanical and semi-classical approaches. Sec. IV focuses on GW-induced broadening in RRLs, comparing it to other broadening mechanisms and discussing its observational feasibility. In Sec. V, we consider the impact of the stochastic gravitational-wave background. Finally, in Sec. VI, we summarize our findings and propose potential avenues for future research.

\section{Effects of Gravitational Waves on the Hyperfine Structure}\label{secII}
Consider a hydrogen atom in a gravitational field characterized by the Riemann curvature tensor ${R^\alpha}_{\beta\gamma\delta}$, where Greek indices denote spacetime coordinates. The primary contribution to the gravitational interaction Hamiltonian, $H_I$, arises from tidal interactions and is given by \cite{parker1980, parker1982, wanwieng2023}:
\begin{equation}\label{HR}
    H_I = \frac{1}{2}\mu c^2 R_{j0k0}x^jx^k,
\end{equation}
where $\mu = m_e m_p/(m_e + m_p)$ is the reduced mass of the atom, $c$ is the speed of light in vacuum, the coordinates $x^0$ and $x^j$ indicate time and spatial coordinates in the atom's local inertial frame, respectively. For linearized gravitational waves, the curvature tensor takes the form:
\begin{equation}\label{Rgw}
R_{j0k0} = -\frac{1}{2c^2} \ddot{h}^{\mathrm{TT}}_{jk},
\end{equation}
where $h^{\mathrm{TT}}_{jk}$ is the gravitational-wave tensor in the Transverse-Traceless (TT) gauge, and the double dots denote the second derivative with respect to time $t$.

Assuming the waves propagate in a well-defined direction $\hat{n}$, we can write
\begin{equation}
     h^{\mathrm{TT}}_{jk}(t) = \sum\limits_{A=+,\times} h_A(t)e^A_{jk}(\hat{n}),
\end{equation}
where $h_{+,\times}(t)$ are the waveforms for the plus and cross polarizations, and $e^A_{jk}(\hat{n})$ is the polarization tensor defined by
\begin{subequations}\label{polarizations}
    \begin{align}
   e^+_{jk}(\hat{n}) &=  \rev{\hat{u}_j{\hat{u}}_k -\hat{v}_j{\hat{v}}_k}\\
    e^\times_{jk}(\hat{n}) &=  \rev{\hat{u}_j{\hat{v}}_k +\hat{v}_j{\hat{u}}_k,}
\end{align}
\end{subequations}
where $\mathbf{\hat{u}},\mathbf{\hat{v}}$ are unit vectors orthogonal to $\hat{n}$ and to each other. We align the direction of gravitational-wave propagation along the $z$-axis, so that the components of $h^{\mathrm{TT}}_{jk}$ take the form
\begin{equation}
    (h^{\mathrm{TT}}_{jk}) = \left( {\begin{array}{*{20}{c}}
h_+(t)&h_\times(t)&0\\
h_{\times}(t)&-h_+(t)&0\\
0&0&0
\end{array}} \right).
\end{equation}
From Eq. \eqref{Rgw} and \eqref{HR}, the gravitational-wave tidal interaction Hamiltonian is
\begin{equation}\label{Hgw}
    H_{\mathrm{gw}} = \frac{1}{4}\mu \ddot{h}^{\mathrm{TT}}_{jk}x^jx^k.
\end{equation}
\rev{
The objective of this section is to investigate the hyperfine structure of the hydrogen atom in the presence of gravitational waves. Specifically, we examine how gravitational waves influence the interaction between the atomic electron and the magnetic dipole moment of the proton, which gives rise to hyperfine splitting.
In flat spacetime, the hyperfine interaction Hamiltonian is given by}
\begin{equation} 
\rev{H^0_{\mathrm{hf}} = 2\frac{\mu_B}{\hbar} \delta_{ij} S^i B^j_0,}
\end{equation}
\rev{where the superscript zero indicates the flat-spacetime expression, $\mu_B = e\hbar/(2m_e)$ is the Bohr magneton, $S^i$ is the electron spin operator, and $B^j_0$ is the magnetic field produced by the proton's magnetic dipole moment $\vec{\mu}_I = \gamma_p \vec{I} = (g_p e/2m_p)\, \vec{I}$. Here, $\vec{I}$ is the proton spin operator and $g_p$ is the proton's $g$-factor. In the absence of gravitational waves, the magnetic field from the proton is} 
\begin{equation}\label{BnoGW}
   \rev{ \vec{B}_0 = \frac{\mu_0 \gamma_p}{4\pi} \frac{3(\vec{I} \cdot \hat{x})\hat{x} - \vec{I}}{r^3} + \frac{2\mu_0 \gamma_p}{3} \vec{I} \delta^3(\vec{x}),}
\end{equation}
\rev{where $\hat{x} \equiv \vec{x}/|\vec{x}|$ and $r \equiv |\vec{x}|$ is the distance between the proton and the electron. This interaction leads to the well-known splitting of the hydrogen ground state $1s^2\, \mathrm{S}_{1/2}$ into two hyperfine levels.}

\rev{When gravitational-wave effects are taken into account, the expression for the magnetic field in Eq.~\eqref{BnoGW} must be modified to satisfy Maxwell's equations in curved spacetime. Moreover, the curvature of spacetime introduces a \textit{direct gravitational coupling} between the electron spin and the background geometry. As shown in Refs.~\cite{quach2016,ito2020}, this modifies the spin-magnetic interaction Hamiltonian to}
\begin{equation}\label{Hspin_full}
    \rev{H_{\mathrm{spin}} = \frac{\mu_B}{\hbar} (2\delta_{ij} + h_{ij}) S^i B^j,}
\end{equation}
\rev{where $h_{ij}$ is the spatial metric perturbation representing the gravitational wave.}
\rev{The hyperfine Hamiltonian in the presence of a gravitational wave thus consists of three contributions:}
\begin{equation}\label{Hhf_total}
    \rev{H_{\mathrm{hf}} = H^0_{\mathrm{hf}} + \Delta H_1 + \Delta H_2,}
\end{equation}
\rev{where the correction due to the gravitationally modified magnetic field is}
\begin{equation}
   \rev{ \Delta H_1 = 2 \frac{\mu_B}{\hbar} \delta_{ij} S^i \delta B^j,}
\end{equation}
\rev{and the direct spin-curvature coupling contribution is}
\begin{equation} \label{Hspin_term}
   \rev{ \Delta H_2 = \frac{\mu_B}{\hbar} h_{ij} S^i B^j_0.}
\end{equation}
\rev{Since the tidal interaction described by $H_{\mathrm{gw}}$ in Eq.~\eqref{Hgw} does not affect the ground-state $1\mathrm{S}$ energy level, as shown in Ref.~\cite{wanwieng2023}, the only gravitational-wave contributions to the hyperfine structure are those contained in $\Delta H_1 + \Delta H_2$.}

\subsection{Electrodynamics in gravitational-wave background}
In quantum mechanics, electric and magnetic fields yield different effects on hydrogen spectra. To calculate such effects, one needs to make distinction between electric and magnetic fields in curved spacetime, which depends on the distinction between space and time in the chosen frame of reference. To do so, we employ the ADM formalism, which is decomposing the 4-dimensional spacetime into a stack of non-intersecting spatial hypersurfaces, $\Sigma_t$, labeled by a global time coordinate $t$. On each hypersurface, spatial coordinates $x^j$ are introduced. The metric in this $3+1$ decomposition takes the form
\begin{equation}
    ds^2 = -\alpha^2 (cdt)^2 + \gamma_{ij}(dx^i + \beta^i cdt)(dx^j + \beta^jcdt) \label{ADMmetric},
\end{equation}
where $\alpha$ is the lapse function, $\beta_i$ is the shift vector, and $\gamma_{ij}$ is the induced metric on the spatial hypersurface $\Sigma_t$. From this, we can extract the components of the metric tensor, given by
\begin{equation}\label{ADMg}
    (g_{\mu\nu}) = \left(\begin{array}{*{20}{c}}
\displaystyle -\alpha^2 +  \beta_k \beta^k& \beta_i\\
\displaystyle \beta_i &\gamma_{ij}
\end{array}\right).
\end{equation}
Here, $\beta_i = \gamma_{ij}\beta^j$, and $\beta^\mu = (0, \beta^i)$. Additionally, we have $\gamma^{00} = \gamma^{0i} = 0$.

The inverse metric, $g^{\mu\nu}$, is given by
\begin{equation}\label{ADMginv}
    (g^{\mu\nu}) = \left(\begin{array}{*{20}{c}}
\displaystyle-\alpha^{-2}&\displaystyle \alpha^{-2}\beta^i\\
\displaystyle\alpha^{-2}\beta^i&\displaystyle \gamma^{ij} - \alpha^{-2}\beta^i\beta^j
\end{array}\right),
\end{equation}
where $\gamma^{ij}$ is the inverse of $\gamma_{ij}$. It is important to note that $g_{ij} = \gamma_{ij}$, but $g^{ij} \neq \gamma^{ij}$ unless $\beta_j = 0$.

The determinant of the metric tensor, $g_{\mu\nu}$, is
\begin{equation}
    g \equiv \mathrm{det}\;g_{\mu\nu} = -\alpha^2 \gamma,
\end{equation}
where $\gamma$ is the determinant of the induced metric $\gamma_{ij}$. This decomposition allows us to study the dynamics of spacetime in a convenient way, especially when considering perturbations such as gravitational waves.

For Eulerian observers, whose timelike worldlines are orthogonal to $\Sigma_t$, their 4-velocity components are expressed as
\begin{equation}\label{obvel}
    u_0 = -\alpha,\quad u_i = 0, \quad u^0 = \frac{1}{\alpha},\quad u^i = -\frac{\beta^i}{\alpha}.
\end{equation}
The induced metric in 4-tensor form can be written as
\begin{equation}
    \gamma^{\mu\nu} = g^{\mu\nu} + u^\mu u^\nu , 
\end{equation}
which ensures that $\gamma^{\mu\nu} u_\mu = 0$, making $\gamma^{ij}$ purely spatial.

In curved spacetime, the electromagnetic fields are described by the antisymmetric field tensor $F^{\mu\nu}$, which satisfies Maxwell's equations:
\begin{subequations}\label{Max1}
\begin{equation}\label{inhomoMax}
    \nabla_\nu F^{\mu\nu} = \mu_0 J^\mu
\end{equation}
and 
\begin{equation}\label{homoMax}
    \epsilon^{\alpha\beta\gamma\delta}\nabla_\beta F_{\gamma\delta} = 0,
\end{equation}
\end{subequations}
where $J^\mu$ is charge-current 4-vector. 

Maxwell's equations in the $3+1$ formalism are derived in terms of physical quantities measured by observers with 4-velocity $u^\mu$, such as the electric field $E^j$, magnetic field $B^j$, charge density $\rho_e$, and current density $j^k$. These fields are regarded as 4-vectors $E^\mu$ and $B^\mu$ on $\Sigma_t$, defined by 
\begin{subequations}
    \begin{align}\label{EBu}
    E^\mu &= F^{\mu\nu}u_\nu,\quad E^\mu u_\mu = 0,\\
    B^\mu &= -\frac{1}{2}\epsilon^{\mu\rho\sigma\nu}u_\rho F_{\sigma\nu},\quad B^\mu u_\mu =0.
\end{align}
\end{subequations}
The Levi-Civita tensor $\epsilon^{\alpha\beta\gamma\delta}$ is given by
\begin{equation}
    \epsilon_{\alpha\beta\gamma\delta} = \sqrt{-g}[\alpha\beta\gamma\delta],\quad \epsilon^{\alpha\beta\gamma\delta} = -\frac{1}{\sqrt{-g}}[\alpha\beta\gamma\delta] 
\end{equation}
where $[\alpha\beta\gamma\delta]$ denotes the permutation symbol. The charge and current densities are given by
\begin{subequations}
    \begin{align}
        \rho_e &= - J^\mu u_\mu,\\
    j^\mu &= \gamma^{\mu\nu}J_\nu.
    \end{align}
\end{subequations}
Conversely, the electromagnetic field tensor and charge-current density can be expressed in terms of the frame's 4-velocity as 
\begin{subequations}\label{FJ}
    \begin{align}
    F^{\mu\nu} &= u^\mu(E^\nu/c) - u^\nu(E^\mu/c) + \epsilon^{\mu\nu\rho\sigma}u_\rho B_\sigma\\
    J^\mu &= \rho_e u^\mu + j^\mu
\end{align}
\end{subequations}

Substituting these into Maxwell's equations yields the $3+1$ formalism of Maxwell's equations \cite{thorne1982}:
\begin{subequations}\label{Max3plus1}
\begin{align}
^{(3)}\nabla_k E^k &= \frac{\rho_e}{\epsilon_0}\\
^{(3)}\nabla_k B^k &= 0\\
D_{u} E^i + \frac{2}{3}\theta E^i - \gamma^{jk}{\sigma^i}_j E_k &= -\frac{1}{\alpha}\epsilon^{ijk} 
 {^{(3)}}\nabla_j\left(\alpha B_k \right) - \mu_0 j^i\\
 D_{u} B^i + \frac{2}{3}\theta B^i - \gamma^{jk}{\sigma^i}_j B_k &= -\frac{1}{\alpha}\epsilon^{ijk} 
 {^{(3)}}\nabla_j\left(\alpha E_k \right).
\end{align}
\end{subequations}
Here, $^{(3)}\nabla_k$ denotes the covariant derivative with respect to the induced metric $\gamma_{ij}$ on $\Sigma_t$. The Fermi time derivative of $A^\mu$ is defined as $D_u A^\alpha \equiv u^\mu \nabla_\mu A^\alpha - u^\alpha a_\mu A^\mu$, where $a_\mu = u^\nu \nabla_\nu u_\mu$ represents the frame's 4-acceleration. The Levi-Civita tensor on $\Sigma_t$ is $\epsilon^{ijk} = \gamma^{-1/2} [ijk]$. The expressions for $\theta$ and $\sigma_{\mu\nu}$ are 
\begin{align}
    \theta &= \nabla_\mu u^\mu\\
    \sigma_{\mu\nu} &= \frac{1}{2}{\gamma_\mu}^{\rho}{\gamma_\nu}^{\beta}\left(\nabla_{\beta}u_\alpha + \nabla_{\alpha}u_\beta \right) - \frac{1}{3}\theta \gamma_{\mu\nu}.
\end{align}

In the weak-field gravity approximation, the spacetime metric is expressed as 
\begin{equation}\label{weakg}
    g_{\mu\nu} = \eta_{\mu\nu} + h_{\mu\nu},\quad |h_{\mu\nu}| \ll 1,
\end{equation}
where $\eta_{\mu\nu} = \mathrm{diag}(-1, 1, 1, 1)$ represents the flat Minkowski metric, and $h_{\mu\nu}$ is the perturbation. To first order in $h_{\mu\nu}$, the inverse metric is 
\begin{equation}\label{weakginv}
    g^{\mu\nu} = \eta^{\mu\nu} - h^{\mu\nu}.
\end{equation}
The Christoffel connections are given by
\begin{equation}
    {\Gamma^\mu}_{\rho\sigma} = \frac{1}{2}({h^\mu}_{\rho,\sigma} + {h^\mu}_{\sigma,\rho} - {h_{\rho\sigma}}^{,\mu}). 
\end{equation}

To express the weak-field metric in ADM form, we use Eq. \eqref{ADMg}, \eqref{ADMginv}, \eqref{weakg}, and \eqref{weakginv} to identify the lapse function and shift vector 
\begin{align}\label{lapseshift}
    \alpha &= 1 - \frac{1}{2}h^{00},\quad \beta_i = h_{i0},\\
    \beta^i &= - h^{i0}, \quad \gamma_{ij} = \delta_{ij} + h_{ij},\quad \gamma^{ij} = \delta^{ij} - h^{ij}. 
\end{align}
The metric determinant is 
\begin{equation}\label{det}
    g = - (1 + \eta^{\mu\nu}h_{\mu\nu}) = -(1 - h_{00} + \delta^{ij}h_{ij}).
\end{equation}
For the normal observer, Eq. \eqref{obvel} gives
\begin{equation}
    u^0 = 1 + \frac{1}{2}h^{00},\quad u^i = h^{i0},\quad
    u_0 = - \left(1 - \frac{1}{2}h^{00}\right),\quad u_i = 0.
\end{equation}
These results hold for the linearized theory of GR in an arbitrary gauge.

To analyze the interaction between an atom and GWs, we adopt a proper reference frame co-moving along the atom's geodesic. In the vicinity of this geodesic, we introduce Fermi normal coordinates (FNC) $x^\mu = (ct,x^k)$, where $t$ is the proper time parameterized the  geodesic and $x^k$ are Cartesian spatial coordinates. The spacetime metric in FNC can be expanded up to quadratic order of $x^j$, as follows \cite{manasse1963}
\begin{subequations}
    \begin{align}
    g_{00} &= -1 - R_{0l0m}(t)x^lx^m \\
    g_{0i} &= -\frac{2}{3}R_{0lim}(t)x^lx^m\\
    g_{ij} &= \rev{\delta_{ij} } -\frac{1}{3}R_{iljm}(t)x^lx^m\\
    g^{00} &= -1 + R_{0l0m}(t)x^lx^m\\
    g^{0i} &= g_{0i}\\
    g^{ij} &= \delta^{ij} + \frac{1}{3}{{{R^i}_l}^j}_m (t)x^lx^m  ,
\end{align}
\end{subequations}
where the Riemann tensor is evaluated at spatial origin, i.e. along the atom geodesic, so it depends only on time $t$. For computational convenience, we choose the transverse-traceless (TT) gauge, so the Riemannian tensor takes the form \eqref{Rgw}. Consequently, we may redefine the above metric coefficients as
\begin{subequations}\label{gFNC}
    \begin{eqnarray}
    g_{00} = -1 + 2f,\quad g^{00} = -1 - 2f \\
    g_{0i} = g^{0i} = 0,\quad g_{ij} = g^{ij} = \delta_{ij}
\end{eqnarray}
\end{subequations}
where we have defined 
\begin{equation}
    f \equiv \frac{1}{4c^2}\ddot h^{\mathrm{TT}}_{lm}(t)x^lx^m
\end{equation}
From Eq. \eqref{gFNC}, \eqref{lapseshift},\eqref{weakg}, \eqref{weakginv}, one can deduce that the corresponding lapse, shift, and induced metric in the Fermi normal coordinates of the proper frame are given by:
\begin{equation}
    \alpha = 1 - f,\quad \beta_i = 0,\quad \gamma_{ij} = \delta_{ij}.
\end{equation}
The frame's 4-velocity is thus
\begin{align}
    u^0 = 1 + f,\quad u_0 &= - 1 +  f,\quad u_i = u^i = 0.
\end{align}
The Maxwell's equations \eqref{Max3plus1} become
\begin{subequations}\label{MWGW}
\begin{align}
    \partial_k E^k  &= \frac{\rho_e}{\epsilon_0}\\
\partial_k B^k  &= 0 \\ 
\frac{\partial B^j}{\partial t} + \epsilon_{jkl}\partial_k[(1 - f)E_l] &=  0 \\
\epsilon_{jkl}\partial_k [(1 -  f )B_l] - \frac{\partial E^j}{\partial t} &= \mu_0 j^j,
\end{align}
\end{subequations}
which describe the electromagnetic fields in the presence of linearized GWs, consistent with the \cite{hwang2023}.

\subsection{The Hyperfine Structure in the Presence of Gravitational Waves}
\rev{The magnetic field produced by a point-like magnetic dipole moment $\vec{\mu}$, satisfying Maxwell’s equations in the presence of GWs, is given by Eq.~\eqref{MWGW}:}
\begin{equation}\label{BGW}
    \rev{\vec{B} = \frac{\mu_0}{4\pi} \frac{3(\vec{\mu} \cdot \hat{x}) \hat{x} - \vec{\mu}}{r^3}
    \left(1 - \frac{1}{4c^2} \ddot{h}^{\mathrm{TT}}_{jk} x^j x^k \right)
    + \frac{2\mu_0}{3} \vec{\mu} \delta^3(\vec{x}).}
\end{equation}
\rev{Comparing Eq.~\eqref{BGW} with the flat-spacetime expression in Eq.~\eqref{BnoGW}, one can identify the GW-induced correction $\delta B^j$. This leads to a perturbation in the hyperfine interaction of the form}
\begin{equation}
    \rev{\Delta H_1} = -K\, \ddot{h}^{\mathrm{TT}}_{lm} \frac{x^l x^m}{r^3}
    \left[3(\vec{I} \cdot \hat{x})(\vec{S} \cdot \hat{x}) - \vec{I} \cdot \vec{S} \right],
\end{equation}
\rev{where the coupling constant is defined as}
\begin{equation}
    \rev{K \equiv \frac{g_p \mu_0 e^2}{32\pi m_e m_p c^2}.}
\end{equation}

\rev{The direct spin-curvature coupling contribution, introduced in Eq.~\eqref{Hspin_term}, takes the form}
\begin{equation}\label{DH2}
   \rev{ \Delta H_2 = - \frac{\mu_0 \mu_B \gamma_p}{4\pi}
    \frac{3(\vec{I} \cdot \hat{x}) h_{ij} S^i \hat{x}^j - h_{ij} S^i I^j}{r^3}.}
\end{equation}

\rev{For concreteness, we consider a monochromatic plane gravitational wave with plus polarization and angular frequency $\omega_{\mathrm{gw}}$ propagating along the $z$-axis. In spherical coordinates, the relevant contraction simplifies to}
\begin{equation}
    \rev{\ddot{h}^{\mathrm{TT}}_{jk} x^j x^k = \omega_{\mathrm{gw}}^2 h_+ r^2 \sin^2\theta \cos 2\phi.}
\end{equation}

\rev{A direct calculation shows that \( h_{ij} S^i \hat{x}^j = 0 \), eliminating the first term in Eq.~\eqref{DH2}. The remaining contribution then reduces to}
\begin{equation}
    \rev{\Delta H_2 = C \frac{h_{ij} S^i I^j}{r^3}, \quad
    C \equiv \frac{\mu_0 g_p e^2}{16\pi m_e m_p}.}
\end{equation}
\rev{The unperturbed hyperfine Hamiltonian $H^0_{\mathrm{hf}}$ is diagonal in the coupled spin basis states $|F M_F\rangle_g$, where the subscript $g$ denotes the ground electronic state with quantum numbers $n = 1$, $\ell = 0$, and $m_\ell = 0$. The matrix elements are}
\begin{subequations}
\begin{align}
    \rev{{}_g\langle 00|H^0_{\mathrm{hf}}|00\rangle_g} &= -\frac{3}{4}A, \\
    \rev{{}_g\langle 1M'_F|H^0_{\mathrm{hf}}|1M_F\rangle_g} &= \frac{1}{4}A\, \delta_{M'_F M_F},
\end{align}
\end{subequations}
where the hyperfine structure constant $A$ is defined as
\begin{equation}
    A = \frac{4g_p \hbar^4}{3 m_p m_e^2 c^2 a^4}.
\end{equation}

\rev{The gravitational wave introduces additional perturbations with matrix elements}
\begin{align}
    \rev{{}_g\langle F' M'_F|\Delta H_1|F M_F\rangle_g} 
    &= \frac{8\pi K}{5a} \omega_{\mathrm{gw}}^2 h_+ \langle F' M'_F| \mathcal{S}_+ |F M_F\rangle, \\
    {}_g\langle F' M'_F| \Delta H_2 |F M_F\rangle_g
    &= \frac{C}{18 a c^2} \omega_{\mathrm{gw}}^2 h_+ \langle F' M'_F| \mathcal{S}_+ |F M_F\rangle,
\end{align}
\rev{where $\mathcal{S}_+ \equiv I_x S_x - I_y S_y$.}

To evaluate the matrix elements of $\mathcal{S}_+$, we expand the coupled spin states $|F M_F\rangle$ in terms of the uncoupled basis $|s_1 m_1; s_2 m_2\rangle$, where $s_1 = s_2 = 1/2$. Using the notation $|\frac{1}{2}, \pm\frac{1}{2}; \frac{1}{2}, \pm\frac{1}{2}\rangle \equiv |\pm\,\pm\rangle$, the states are
\begin{subequations}
\begin{align}
    |00\rangle &= \frac{1}{\sqrt{2}} (|+\,-\rangle - |-\,+\rangle), \\
    |11\rangle &= |+\,+\rangle, \\
    |10\rangle &= \frac{1}{\sqrt{2}} (|+\,-\rangle + |-\,+\rangle), \\
    |1{-}1\rangle &= |-\,-\rangle.
\end{align}
\end{subequations}

A straightforward computation yields the only non-zero matrix elements:
\begin{equation}
    \langle 1{-}1|\mathcal{S}_+|11\rangle = \langle 11|\mathcal{S}_+|1{-}1\rangle = \frac{\hbar^2}{2}.
\end{equation}

\rev{Combining all contributions, the hyperfine Hamiltonian matrix in the coupled basis is}
\begin{equation}
\rev{H_{\mathrm{hf}} =
\begin{pmatrix}
    -\frac{3}{4}A & 0 & 0 & 0 \\
    0 & \frac{1}{4}A & 0 & \Delta_{\mathrm{gw}} \\
    0 & 0 & \frac{1}{4}A & 0 \\
    0 & \Delta_{\mathrm{gw}} & 0 & \frac{1}{4}A
\end{pmatrix},}
\end{equation}
\rev{where}
\begin{equation}
   \rev{ \Delta_{\mathrm{gw}} \equiv \frac{\mu_0 g_p e^2 \hbar^2}{576 \pi m_e m_p a c^2}
    \left( \frac{72\pi}{5} + 1 \right) \omega_{\mathrm{gw}}^2 h_+.}
\end{equation}

\begin{figure}[htb]
    \centering
    \includegraphics[width=.7\columnwidth]{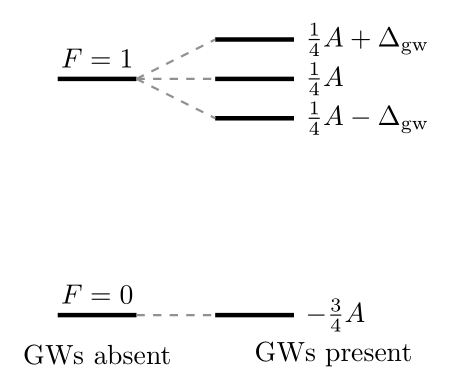}
    \caption{Hyperfine structure of $1s ^{2}\mathrm{S}_{1/2}$ hydrogen in the absence and presence of gravitational waves.}
    \label{fig:hyperfine}
\end{figure}

\rev{Diagonalizing this matrix yields the energy eigenvalues}
\begin{align}
    \rev{E_{F=-} = -\frac{3}{4}A,\quad E_{F=1}=\frac{1}{4}A,\quad \frac{1}{4}A \pm \Delta_{\mathrm{gw}}.}
\end{align}
The corresponding energy level diagram is shown in Fig.~\ref{fig:hyperfine}. The fractional shift of the hyperfine splitting induced by the gravitational wave is
\begin{equation}
    \frac{\Delta \nu}{\nu} = \frac{2\Delta_{\mathrm{gw}}}{A} \simeq 5.9 \times 10^{-35}
    \left( \frac{f_{\mathrm{gw}}}{\mathrm{MHz}} \right)^2
    \left( \frac{h_+}{10^{-9}} \right),
\end{equation}
indicating that the splitting induced by gravitational waves is extremely small and well below the sensitivity of current experimental techniques.

\section{Effect of Gravitational Waves on Rydberg States}\label{secIII}
Another radiative process responsible for radio-frequency lines in hydrogen is the transition between states with high principal quantum numbers, i.e., Rydberg states. In this regime, the effects of GWs on both fine and hyperfine structure can be neglected, as the corresponding interactions weaken with increasing electron-nucleus separation, whereas the gravitational tidal interaction scales quadratically with distance. Thus, the relevant Hamiltonian is
\begin{equation}\label{HRRL}
    H = H_0 + H_{\mathrm{gw}},
\end{equation}
where 
\begin{equation}\label{H0}
    H_0 = \frac{p^2}{2\mu} - \frac{k}{r},\quad k = \frac{e^2}{4\pi\epsilon_0},
\end{equation}
and $H_{\mathrm{gw}}$ is given in Eq, \eqref{Hgw}.

\subsection{Quantum mechanical approach}
In the quantum-mechanical framework of perturbation theory, the interaction Hamiltonian $H_{\mathrm{gw}}$ is introduced as a small perturbation to the Bohr Hamiltonian $H_0$, consistent with the weak nature of gravitational waves. Given the significant degeneracy in hydrogen's energy levels, first-order degenerate perturbation theory is used to determine the associated energy shifts.

\begin{figure}
    \centering
    \includegraphics[width=0.9\linewidth]{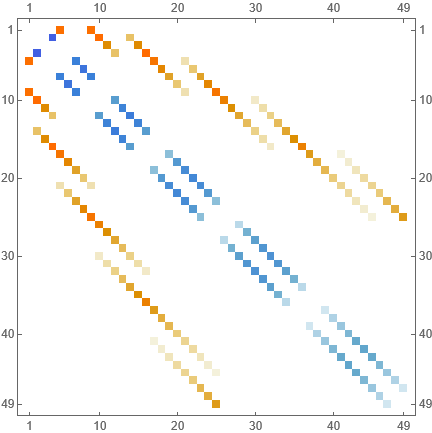}
    \caption{Matrix representation of gravitational-wave perturbation in the energy level $n=7$ of hydrogen, which exhibits $49$-fold degeneracy. The structure is sparse and banded, with distinct groupings of nonzero elements, reflecting selection rules imposed by gravitational-wave coupling.}
    \label{fig1:matrix}
\end{figure}

Expressed in spherical coordinates ($r,\theta,\phi$), the perturbation Hamiltonian from Eq. \eqref{Hgw} takes the form
\begin{equation}\label{Hqm}
    H_{\mathrm{gw}} = \frac{1}{4}\mu\omega^2_{\mathrm{gw}}r^2\sin^2\theta(h_+\cos 2\phi  + h_\times \sin 2\phi ),
\end{equation}
where the angular dependence can be expanded in terms of spherical harmonics $Y^m_\ell(\theta,\phi)$. Consider only the term containing $h_+$, the first-order energy shifts are obtained by computing the eigenvalues of the perturbation matrix
\begin{equation}\label{element}
    \langle n \ell m|H^+_{\mathrm{gw}} |n \ell'm' \rangle = \kappa  \mathcal{R}^{n\ell}_{n\ell'}\mathcal{F}^{mm'}_{\ell\ell'}
\end{equation}
where, $\kappa \equiv \mu\omega^2_{\mathrm{gw}}h_+/4$. The radial and angular components are defined as
\begin{equation}
    \mathcal{R}^{n\ell}_{n\ell'} \equiv \langle n\ell |r^2|n \ell'\rangle
\end{equation}
and
\begin{equation}\label{angmatrix}
 \mathcal{F}^{mm'}_{\ell\ell'} \equiv \sqrt{\frac{8\pi}{15}}\langle \ell m|(Y^2_2 + Y^{-2}_2)|\ell' m' \rangle. 
\end{equation}
For a given principal quantum number $n$, the perturbation matrix has dimensions $n^2\times n^2$, therefore, numerically solving for eigenvalues requires increasingly high computation power for large $n$. Hence, deriving analytical expressions for the matrix elements in Eq. \eqref{element} becomes particularly useful at high $n$. The resulting perturbation matrix for $n=7$ is shown in Fig.~\ref{fig1:matrix}, which highlights its sparse and banded structure dictated by the selection rules.

\begin{figure}
    \centering
    \includegraphics[width=\linewidth]{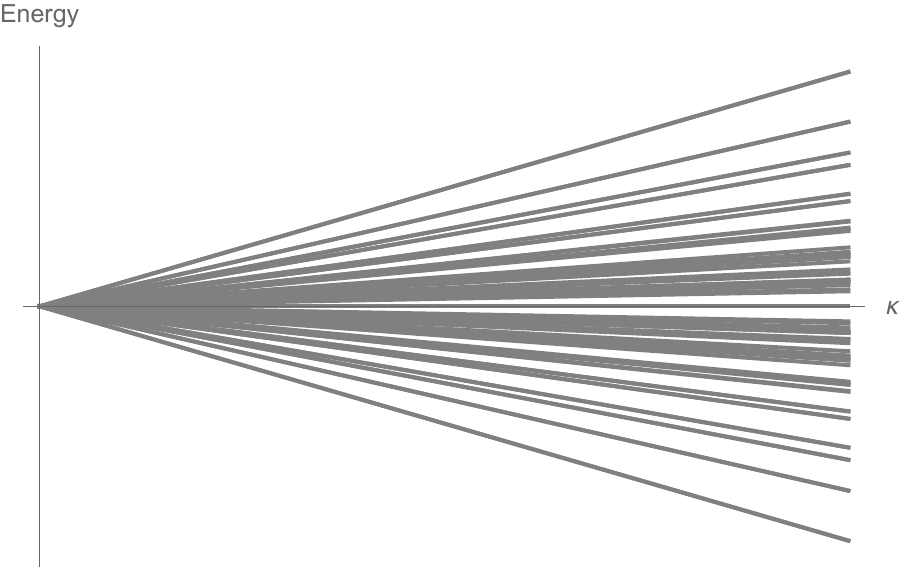}
    \caption{Splitting of the $n=9$ hydrogen energy level due to gravitational-wave perturbation. GWs lift the degeneracy, distributing the energy levels according to selection rules. Each line represents a sublevel’s shift under the perturbation, with $\kappa$ indicating the perturbation strength.}
    \label{fig2:splitting}
\end{figure}

The selection rules for the perturbation matrix elements $\langle n\ell m| H_{\mathrm{gw}}|n \ell'm'\rangle$, as established in Ref. \cite{wanwieng2023}, dictate that nonzero values occur only when
\begin{subequations}
    \begin{equation}
    \Delta\ell \equiv \ell' - \ell = 0,\;\pm 2
\end{equation}
and
\begin{equation}
    \Delta m \equiv m' - m = \pm 2.
\end{equation}
\end{subequations}
For the nonzero elements, we derive explicit analytical formulas for the relevant radial and angular components using the methods detailed in Appendix \ref{appendixA} and \ref{appendixB}. However, since obtaining an exact analytical form of the eigenvalues, is not practical with this approach, we introduce a semi-classical treatment in the next subsection. The resulting lifting of degeneracy for $n=9$ levels is illustrated in Fig.~\ref{fig2:splitting}.

\subsection{Semi-classical approach}
In the classical hydrogen atom model, the electron's motion within the Coulomb potential follows a Keplerian orbit, with a total energy $E < 0$. The corresponding Hamiltonian, $H_0$, is given by Eq. \eqref{H0}. Both the total energy $E$ and the orbital angular momentum $L$ are conserved, reflecting the static and isotropic nature of the Coulomb potential.

\begin{figure}
    \centering
    \includegraphics[width=\linewidth]{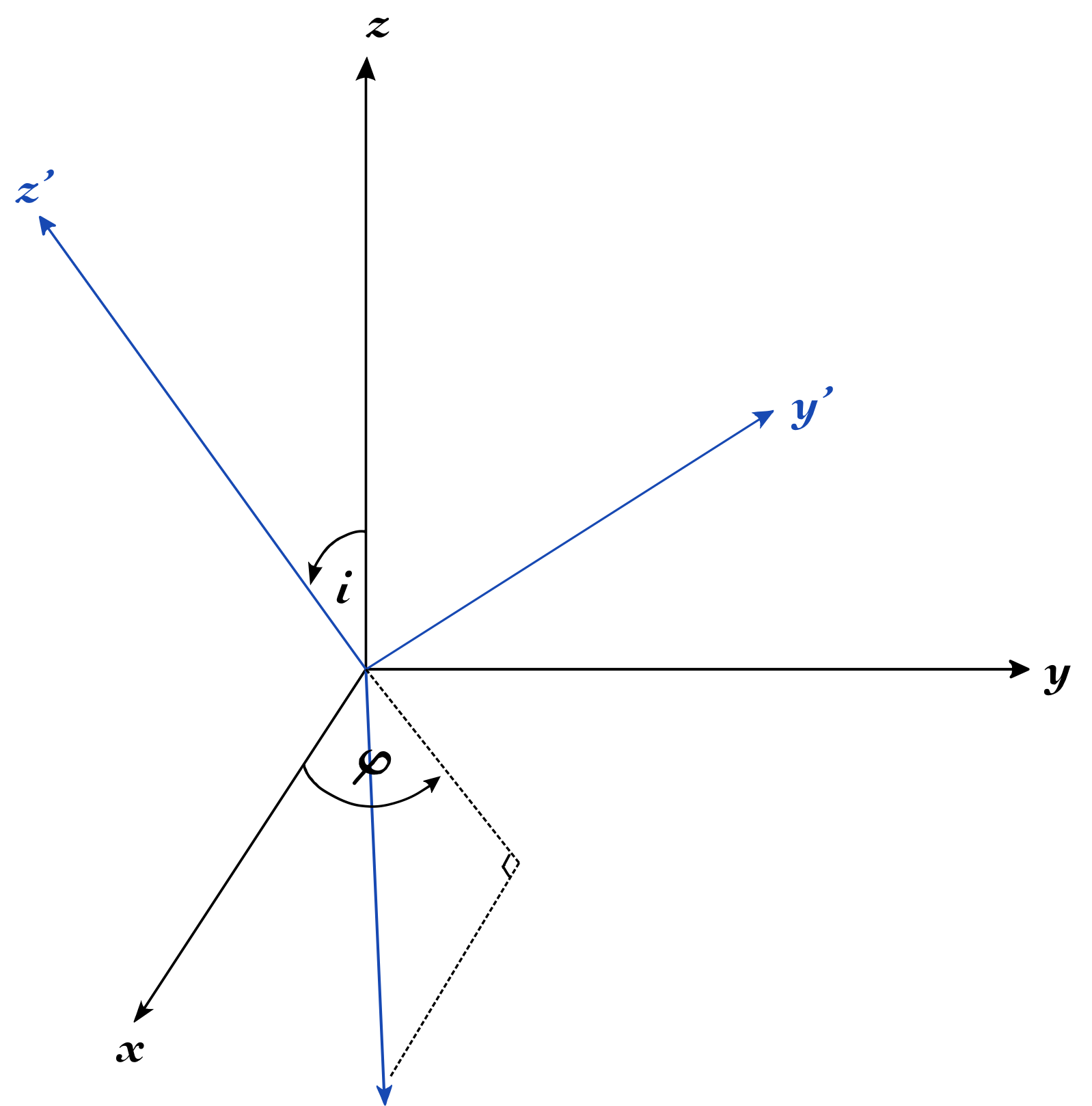}
    \caption{The transformation from the original coordinate system ($x,y,z$) to the rotated frame ($x',y',z'$), where the electron's orbit lies in the $x'y'$-plane.}
    \label{rotatedframe}
\end{figure}

When a gravitational wave is present, the interaction between the atom and the wave, within the Newtonian framework, is described by $H_{\mathrm{gw}} = \mu\Phi_{\mathrm{gw}}$, where the gravitational potential $\Phi_{\mathrm{gw}} = -(1/4)\omega^2_{\mathrm{gw}}h^{\mathrm{TT}}_{jk}x^jx^k$. To calculate this energy shift, we use classical perturbation theory i.e. averaging the perturbation Hamiltonian over the unperturbed motion of the electron. Specifically, for this system, the perturbation Hamiltonian $H_{\mathrm{gw}}$ is averaged over one orbital period $\tau_0$ of the electron's Keplerian motion, denoted as `$\langle\ldots\rangle_0$'', 
\begin{equation}\label{effectiveHgw}
    \delta E_{\mathrm{gw}} = \left\langle H_{\mathrm{gw}} \right\rangle_0 \equiv \frac{1}{\tau_0}\int_0^{\tau_0}H_{\mathrm{gw}}dt.
\end{equation}
To simplify the computation of the term $h^{\mathrm{TT}}_{jk} x^{j} x^{k}$, we choose to work in a rotated coordinate frame $(x', y', z')$, where the electron's orbit lies entirely in the $x'y'$-plane. As shown in Fig. \ref{rotatedframe}, this new frame is related to the original frame by two successive rotations: one by an angle $\iota$ around the $y$-axis, followed by another by an angle $\varphi$ around the $z$-axis. This transformation is represented by the rotation matrix $\mathcal{R}_{ij}$:
\begin{equation}
    (\mathcal{R}_{ij}) = \left( {\begin{array}{*{20}{c}} 
\cos\varphi & \sin\varphi & 0\\
-\sin\varphi & \cos\varphi &0\\
0&0&1
\end{array}} \right)\left( {\begin{array}{*{20}{c}}
\cos\iota& 0&\sin\iota \\
0&1&0\\
-\sin\iota &0&\cos\iota
\end{array}} \right).
\end{equation}

Each component $h^\mathrm{TT}_{j'k'}$ in the rotated frame $(x',y',z')$ is given by
\begin{equation}
    h^{\mathrm{TT}}_{i'j'} = \mathcal{R}_{ik}\mathcal{R}_{jl}h^{\mathrm{TT}}_{kl}.
\end{equation}

For ease in the mathematical expressions, and without significantly affecting the physical estimation of the energy shift, we set $\varphi = 0$ for the following calculations. In the rotated coordinate frame, the components of the gravitational wave field are
\begin{equation}
    h_{x'x'} = h_+\cos^2\iota,\quad h_{y'y'} = - h_+,\quad  h_{x'y'} = h_{y'x'} = h_\times \cos\iota.
\end{equation}
The electron's position in this frame can be described using polar coordinates $(r, \psi)$ as
\begin{equation}
      x' = r\cos\psi,\quad y' = r\sin\psi,\quad z' = 0,
\end{equation}
where the radial coordinate $r$ is given by
\begin{equation}\label{Kepradius}
    r = \frac{L^2}{\mu k}[1 + \epsilon\cos \psi]^{-1},
\end{equation}
with $L$ representing the classical orbital angular momentum and $\epsilon$ the orbital eccentricity. Substituting these into Eq. \eqref{Hgw} results in 
\begin{equation}
    H_{\mathrm{gw}} = \frac{\mu\omega^2_{\mathrm{gw}}r^2}{4}\left\{h_+[(\cos^2\iota + 1)\cos\psi - 1] + h_\times \cos\iota \sin2\psi\right\}.
\end{equation}
To evaluate the energy shift, we change variables from $t$ to $\psi$ in the integral from Eq. \eqref{effectiveHgw}, using the relation
\begin{equation}\label{dtdpsi}
    dt = \frac{\mu r^2}{L}d\psi.
\end{equation}
This gives
\begin{equation}\label{Eshiftclassic}
    \delta E_{\mathrm{gw}} = \frac{\mu}{L\tau_0}\int_0^{2\pi}  r^2 H_{\mathrm{gw}} d\psi.
\end{equation}
Under the assumption that the gravitational wave frequency is significantly lower than the electron’s orbital frequency (the adiabatic regime), the time-dependent variation of the GW amplitude can be considered negligible. Consequently, it remains effectively constant over one orbital period. This condition is satisfied when
\begin{equation}\label{adiabatic}
     \omega_{\mathrm{gw}} \ll \frac{\alpha c}{a_0n^3},
\end{equation}
where $\alpha$ is the fine-structure constant.

By substituting \eqref{Kepradius} and \eqref{dtdpsi} into \eqref{Eshiftclassic} and completing the integration, we get
\begin{align}\label{classDelE}
    \delta E_{\mathrm{gw}} = & \frac{\pi \omega^2_{\mathrm{gw}}L^7h_+}{8 k^4 \mu^2\tau_0}\times\notag\\
    &\left[\frac{(2\cos^2 \iota - 1)\epsilon(\epsilon^2 + 4) + 3\epsilon^3 + 6\epsilon^2 + 12\epsilon + 4}{(1 - \epsilon^2)^{7/2}}\right].
\end{align}
Note that only the plus polarization component contributes to the energy shift.

The action variables are introduced as follows \cite{goldstein2002}:
\begin{subequations}\label{actionvar}
    \begin{align}
         I_1 &= L_z, \label{actionvar1}\\
         I_2 &= L, \label{actionvar2}\\
         I_3 &= \sqrt{-\frac{\mu k^2}{2 H_0}}. \label{actionvar3}
    \end{align}
\end{subequations}
These are associated with the angle variables $\phi_1$, $\phi_2$, and $\phi_3$, representing the longitude of the ascending node, the argument of perihelion, and the mean anomaly, respectively. The action variables are constrained by $|I_1| \le I_2 \le I_3$. The orbital inclination $\iota$, eccentricity $\epsilon$, and the period of the Kepler orbit $\tau_0$ can be expressed in terms of the action variables as
\begin{equation}
    \cos\iota = \frac{I_1}{I_2},\quad \epsilon = \sqrt{1 - \left(\frac{I_2}{I_3}\right)^2},\quad \tau_0 = \frac{2\pi I^2_3}{\mu k^2}.
\end{equation}

According to the Wilson-Sommerfeld quantization rule, each classical action variable $I_k$ is quantized as follows:
\begin{equation}
    I_k = \oint p_k dq_k = n_k \hbar
\end{equation}
This leads to the specific quantization conditions
\begin{equation}\label{quantized}
    I_1 = m\hbar,\quad I_2 = \ell \hbar,\quad I_3 = n\hbar,
\end{equation}
 where $m$, $\ell$, and $n$ are integers representing quantum numbers associated with each action variable. Notably, quantizing $I_3$ in Eq.~\eqref{actionvar3} yields the Bohr energy levels for the hydrogen atom:
 \begin{equation}
     H_0 =  - \left(\frac{\mu k^2}{2\hbar^2}\right)\frac{1}{n^2} = E^{\mathrm{Bohr}}_n.
 \end{equation}
By substituting Eq. \eqref{quantized} and \eqref{actionvar} into \eqref{classDelE}, we obtain
\begin{widetext}
    \begin{equation}\label{Eshiftsemi}
    \delta E^{\mathrm{gw}}_{n\ell m} = \frac{5\mu a^2 \omega^2_{\mathrm{gw}} h_+}{8}n^4\left[1 - \frac{3}{5}\left(\frac{\ell}{n}\right)^2 + \left\{1 -\frac{1}{5} \left(\frac{\ell}{n}\right)^2 + \left(\frac{m}{\ell}\right)^2 -\frac{1}{5} \left(\frac{m}{n}\right)^2 \right\}\sqrt{1 - \left(\frac{\ell}{n}\right)^2}\right].
\end{equation}
\end{widetext}
Here, $a = (m_e/\mu)a_0$ and $a_0$ is the Bohr radius. However, this result in Eq. \eqref{Eshiftsemi} becomes problematic as $n \to \infty$ because it diverges, while in reality, the atom would ionize before reaching such high states. Moreover, recall that this result was derived under the perturbative assumption that $H_{\mathrm{gw}} \ll H_0$. Therefore, the approximation remains valid as long as $\delta E^{\mathrm{gw}}_{n\ell m} \ll E^{\mathrm{Bohr}}_n$ and the adiabatic condition in Eq. \eqref{adiabatic} holds.     

\section{Broadening in hydrogen RRLs}\label{secIV}
The radiative transition from $n+1$ to $n$ in the hydrogen atom, referred to as $\mathrm{H}n\alpha$, occurs within the radio-frequency range for $n \gtrsim 100$. The corresponding frequency is approximately given by
\begin{equation}\label{nu0}
    \nu_0 \simeq \frac{k}{a n^3} \approx \frac{6.576\times 10^{15}\;\mathrm{Hz}}{n^3},
\end{equation}
where $a = a_0(m_e/\mu)$ and $a_0$ is the Bohr radius.

As discussed previously, the presence of GWs induces energy shifts in the atomic states, breaking the degeneracy of the Bohr energy level $n$. This results in sublevels symmetrically distributed around the original unperturbed level. As a result, the spectral lines for transitions such as $\mathrm{H}n\alpha$ exhibit broadening, providing a potential signature of GW interactions with atoms. The linewidth broadening of the $\mathrm{H}n\alpha$ transition caused by GWs is quantified by $\Delta\nu = (\delta E^{\mathrm{gw}}_{(n+1)\ell' m'} + \delta E^{\mathrm{gw}}_{n\ell m}) / h$, where $h$ is Planck's constant. 
%

%\NCc{Why would this be related only to $(n+1, n, n)$ and $(n, n-1, n-2)$ and plus instead of minus sign?}
%
Specifically, for high $n$, the semi-classical formula, Eq. \eqref{Eshiftsemi}, and \eqref{nu0} gives
\begin{align}
    \left(\frac{\Delta\nu}{\nu_0}\right)_{\mathrm{gw}} &= \frac{\mu a^3\omega^2_{\mathrm{gw}}h_+ n^7}{2k} \notag\\
    &\simeq 2.53\times 10^{-7} \left(\frac{f_{\mathrm{gw}}}{1\;\mathrm{GHz}}\right)^2\left(\frac{h_c}{10^{-10}}\right)\left(\frac{n}{300}\right)^7.
\end{align}
The spectral resolving power $R$ required to measure the GW-induced linewidth of the $\mathrm{H}n\alpha$ transition for large values of $n$ is given by
\begin{align}
    R &\simeq \left(\frac{\Delta \nu}{\nu _0}\right)_{\mathrm{gw}}^{-1}\\
    &\simeq 5.93\times 10^8\left(\frac{300}{n}\right)^6\left(\frac{1\;\mathrm{GHz}}{f_{\mathrm{gw}}}\right)^2\left(\frac{10^{-10}}{h_+}\right),
\end{align}
and for a given resolving power $R$, the detectable gravitational-wave amplitude is 
\begin{equation}
    h_{\mathrm{det}} = \mathrm{O}\left(\frac{\mu a^3\omega^2_{\mathrm{gw}}n^7}{2kR}\right).
\end{equation}
We can use this quantity to represent the detection sensitivity. 

When comparing different broadening mechanisms, the most fundamental contributions are natural broadening.
\begin{equation} 
\left(\frac{\Delta\nu}{\nu_0}\right)_{\mathrm{rad}}  = \frac{1}{2\pi\nu_0}\Gamma_0,
\end{equation}
where $\Gamma_0$ is the sum of the total rate of spontaneous decay for Rydberg states of principal quantum numbers $n+1$ and $n$, which are approximately equal and given by the sum of Einstein-A coefficients $A_{nn'}$ for $n'< n$:

\begin{equation}
    \Gamma_0 = \Gamma_{n+1} + \Gamma_n \simeq 2\Gamma_n = 2\sum\limits_{n'<n} A_{nn'}.
\end{equation}
In addition, interstellar atoms exist in a finite-temperature environment permeated by blackbody radiation (BBR), such as the cosmic microwave background (CMB). As a result, BBR-induced broadening must be accounted for in the study of Rydberg states \cite{beterov2009,beterov2009c}:
\begin{equation} 
\left(\frac{\Delta\nu}{\nu_0}\right)_{\mathrm{BBR}}  = \frac{1}{2\pi\nu_0}\Gamma_{\mathrm{BBR}},
\end{equation}
 where $\Gamma_{\mathrm{BBR}}$ is the total rate of BBR-induced transitions:
\begin{equation}
    \Gamma_{\mathrm{BBR}} = \sum\limits_{n'} \frac{A_{nn'}}{\exp(\hbar\omega_{nn'}/k_BT) - 1}.
\end{equation}
For analytical convenience, an approximate expression valid for $n > 20$ (see section 5.4.1 of Ref. \cite{sobel2012}) can be used:
\begin{align}
 A_{nn'} &= \frac{2A_0}{n^3n'(n^2 - n'^2)},\\
    \sum\limits_{n'<n} A_{nn'} &= \frac{A_0}{n^5}\ln\left(\frac{n^3 - n}{2}\right)
\end{align}
where $A_0 = 7.89 \times 10^9 \mathrm{s}^{-1}$.

\begin{figure}
    \centering
    \includegraphics[width=\linewidth]{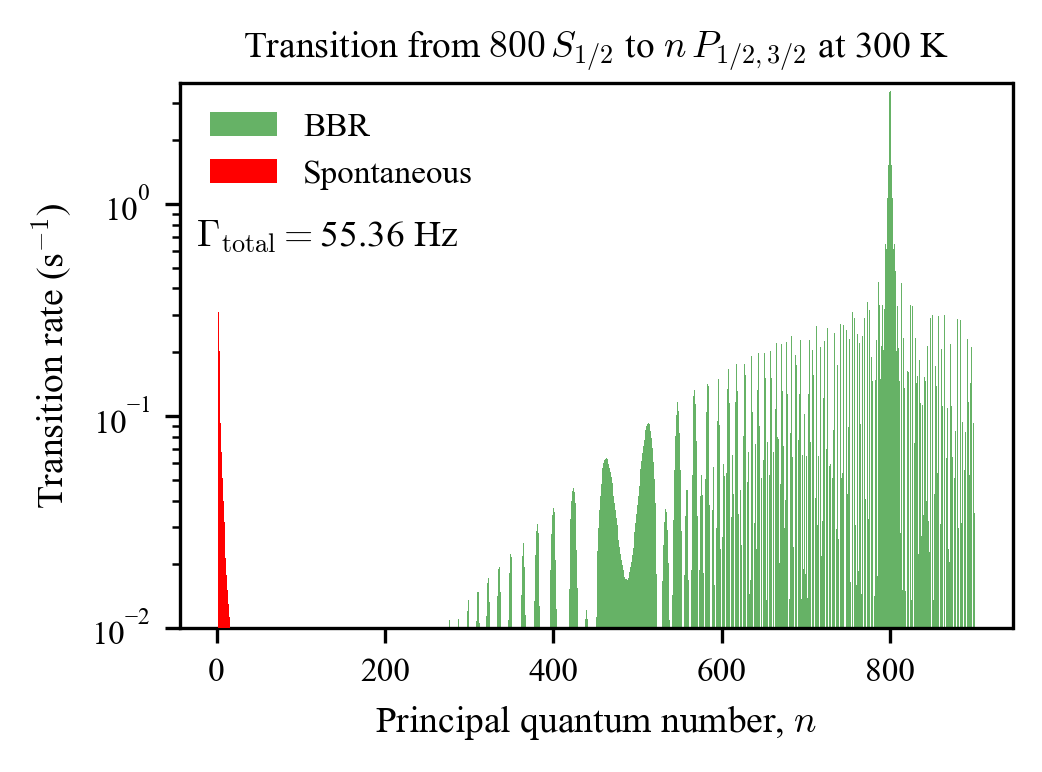}
    \caption{
    Transition rates from hydrogen in the $800S_{1/2}$ state to $nP_{1/2,\,3/2}$ states due to blackbody radiation (BBR) at $T = 300\,\mathrm{K}$ and spontaneous decay. The vertical axis is shown on a logarithmic scale to capture several orders of magnitude in transition rates. The total transition rate is $\Gamma_{\mathrm{total}} = 55.36\,\mathrm{Hz}$.
    }
    \label{BBR300}
\end{figure}

In addition, interstellar atoms exist in various astrophysical environments with different local temperatures and pressures, which influence their atomic spectra and must be taken into account. One of the dominant broadening mechanisms in the observation of RRLs is Doppler broadening. The Doppler linewidth due to the thermal motion of hydrogen gas with kinetic temperature $T$ (in Kelvin) is given by \cite{Gordon2002}
\begin{align}\label{Doppler}
    \left(\frac{\Delta \nu}{\nu_0}\right)_{\mathrm{Doppler}} &= \left(\frac{(8\ln 2) k_BT}{m_\mathrm{H}c^2}\right)^{1/2}\notag\\
    &\simeq (7.137\times 10^{-7}\mathrm{K}^{-1/2})\sqrt{T}.
\end{align}
However, direct calculations using the Alkali Rydberg Calculator (ARC) \cite{vsibalic2017} indicate that for the lowest possible BBR temperature corresponding to the CMB, i.e. $T = 3\,\mathrm{K}$, BBR-induced transitions are negligible. Figure \ref{BBR300} illustrates the transition rates due to BBR at $T=300$ K compared to spontaneous decay for hydrogen at $n=800$.

As a constraint on Doppler broadening, the lowest natural temperature in the universe, in the absence of other heating mechanisms, is set by the cosmic microwave background (CMB) at approximately 2.73 K. Interstellar hydrogen gas can cool to near this temperature through radiative equilibrium with the CMB. In such cases, the Doppler broadening is $1.78\times 10^{-6}$. 

In addition to the aforementioned broadening, another significant effect contributing to spectral line broadening arises from the immersion of hydrogen atoms in a high-density neutral background. Interactions between hydrogen atoms in the Rydberg state and the background gas result in two dominant effects: energy level shifts and spectral line broadening~\cite{Omont1977, Thaicharoen2024}. 
%
%Regarding the spectral line broadening, 
For a Rydberg state with principal quantum number $n$, the resulting line broadening can be expressed as
\begin{equation} % Note that this equation is not in SI unit. We should talk about what unit we want to use.
    \left(\frac{\Delta\nu}{\nu_0}\right)_{\mathrm{sc}}= \frac{a_s N\hbar}{m_ek}an^3 ,
\end{equation}
where $a_{s}$ is the s-wave scattering length of background gas and $N$ is the volume density of the background gas.

Additionally, when hydrogen atoms are in the Rydberg state, the large separation between the electron and proton allows the background gas to occupy the Rydberg-atom volume. The background gas is then subject to the electric field generated by the Rydberg hydrogen atoms. This phenomenon, known as polarization-induced decay, contributes to the additional linewidth in the hydrogen Rydberg spectra,
\begin{equation}
    \left(\frac{\Delta \nu}{\nu_0}\right)_{\mathrm{p}}= \frac{6.21}{2\pi k} \left[\frac{\alpha e^2\hbar^2}{k} \right]^{2/3}an^3v^{1/3} N,
\end{equation}
where $\alpha$ is the polarizability of the background gas and $v$ is the mean relative velocity between the hydrogen atoms and the background gas. In general, $(\Delta\nu/\nu_0)_{\mathrm{p}}$ is larger than $(\Delta\nu/\nu_0)_{\mathrm{sc}}$ and the total linewidth according to these two affect becomes $(\Delta\nu/\nu_0)_{\mathrm{sc}} + (\Delta\nu/\nu_0)_{\mathrm{p}}$.

Note that although we have primarily considered hydrogen so far, this analysis also applies to the RRLs of other species. Any atom in a Rydberg state can be treated approximately as a hydrogen-like system with a different reduced mass.

The characteristic amplitude of quadrupolar gravitational-wave emission from a compact binary system consisting of point-like masses $m_1$ and $m_2$ in quasi-circular Newtonian orbits is given by 
\begin{equation}
    h_c = \frac{4}{D}\left(\frac{G\mathcal{M}}{c^2}\right)^{5/3}\left(\frac{\pi f_{\mathrm{gw}}}{c}\right)^{2/3}
\end{equation}
where $\mathcal{M} = (m_1m_2)^{3/5}/(m_1+m_2)^{1/5}$ is the binary chirp mass, and $D$ represents the distance between the source and observer.

For binary systems made up of compact objects like black holes or neutron stars, the inspiral phase ends when the objects reach the innermost stable circular orbit (ISCO), as determined by Schwarzschild geometry. This occurs when the separation between the objects reduces to $r_{\mathrm{ISCO}} = 6G(m_1 + m_2)/c^2$. After this point, the objects spiral inward, and the strong-field dynamics begin to influence the system's evolution. At this stage, the GW waveform can no longer be described by the linearized theory of general relativity that is assumed here. For compact binary inspirals, the frequency of the GW at ISCO, which marks the peak frequency of the inspiral phase, is given by
\begin{equation}
f_{\mathrm{gw}}^{\mathrm{ISCO}} = \frac{1}{6\sqrt{6}\pi}\frac{c^3}{G M} \simeq 4400\;\mathrm{Hz}\left(\frac{M_{\odot}}{M}\right)
\end{equation}
where $M = m_1+m_2$. The corresponding characteristic amplitude is
\begin{equation}
   h_c^{\mathrm{ISCO}} \simeq  \frac{2}{3}\frac{G}{c^2}\frac{\mu_s}{D} \simeq 6.56\times 10^{-6}\left(\frac{\mu _s}{M_{\odot}}\right)\left(\frac{1\;\mathrm{AU}}{D}\right)
\end{equation}
where $\mu_s = m_1m_2/(m_1+m_2)$.

Binary inspirals of stellar-mass or massive black holes with masses greater than $3M_{\odot}$ produce GW frequencies at ISCO below 100 Hz. To generate GWs with peak frequencies above $10 \, \mathrm{kHz}$ and up to the GHz range, the component masses in the binary system must be on the order of $\sim 10^{-1} - 10^{-6}M_{\odot}$. Such masses are too small for the objects to be stellar-mass black holes or even the smallest neutron stars. Instead, these sub-solar-mass compact objects could be primordial black holes (PBHs) with planetary masses. PBHs are theorized to form in the early Universe through various mechanisms and are considered potential candidates for dark matter (see \cite{sasaki2018, carr2020, bagui2023} for reviews). Unlike stellar-mass black holes, PBHs can have a wide range of masses, from as small as the Planck mass ($10^{-5}\;\mathrm{g}$) to as large as $10^5M_{\odot}$. The GW-induced broadening of RRLs presents a promising observational method for detecting such signals and could provide a new way to confirm the existence of small PBHs formed in the early Universe. As illustrated in Fig.~\ref{fig:linewidth}, the GW contribution to the linewidth can dominate over natural broadening at high $n$ and comparable scales with Doppler effects under certain conditions.

\begin{widetext}
 \begin{center}
        \begin{figure}[h]
    \centering
   \includegraphics[width=.8\textwidth]{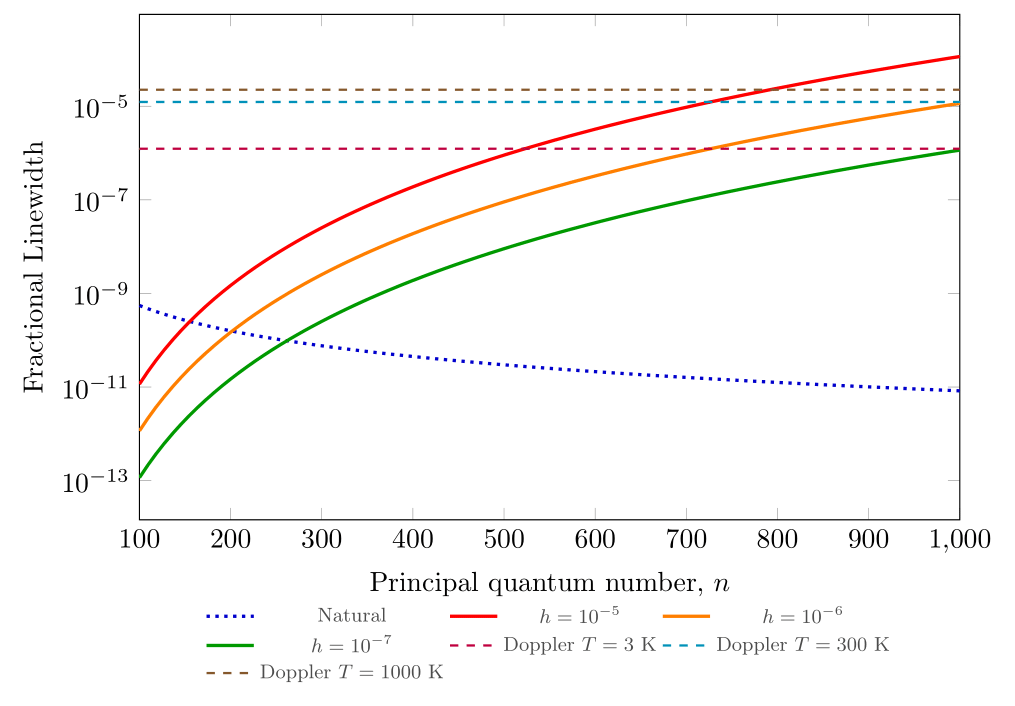}
    \caption{Fractional linewidth contributions from natural broadening, gravitational-wave broadening induced by a MHz-frequency GW from a $10^{-3}M_\odot$ PBH in a binary inspiral at various distances, and Doppler broadening, for hydrogen atoms with principal quantum number $n$.}
    \label{fig:linewidth}
\end{figure}
 \end{center}
\end{widetext}

\section{Stochastic gravitational-wave background}
Previously, we analyzed gravitational waves as nearly monochromatic signals with a well-defined angular frequency 
 $\omega_{\mathrm{gw}}$, traveling in a single direction through atomic systems, typically originating from isolated periodic sources. However, a more general case to consider is the stochastic gravitational-wave background (SGWB), which consists of a random superposition of waves arriving from all directions, covering a broad spectrum of frequencies and amplitudes \cite{maggiore2000}. At the atom’s location $\Vec{x} = 0$, the SGWB is described by
\begin{equation}\label{SGWB}
     h^{\mathrm{TT}}_{jk}(t) = \sum\limits_{A=+,\times}\int d^2\hat{n}\; e^A_{jk}(\hat{n}) \int df\;\tilde{h}_A(f,\hat{n}) e^{-2\pi fti}.
\end{equation}
Here $\hat{n}$ denotes the wave propagation direction, with $d^2\hat{n} = d\cos\theta d\phi$. The Fourier amplitude $\tilde{h}_A(f,\hat{n})$ satisfies the statistical condition
\begin{equation}
    \langle \tilde{h}^*_A(f,\hat{n})\tilde{h}_A(f',\hat{n'})\rangle = \delta_{AA'}\delta(f-f')\frac{\delta^2(\hat{n},\hat{n'})}{4\pi}\frac{1}{2}S_{h}(f),
\end{equation}
 where $\langle \ldots \rangle$ denotes the ensemble average, and $\delta^2(\hat{n},\hat{n'}) = \delta(\phi -\phi')\delta(\cos\theta - \cos\theta')$. The function $S_h(f)$ is the single-sided spectral density of the SGWB, which is related to the energy density per unit logarithmic frequency
\begin{equation}
    \Omega_{\mathrm{gw}}(f) = \frac{1}{\rho_c}\frac{d \rho_{\mathrm{gw}}}{d \log f},
\end{equation}
where $\rho_c$ is the critical density, by the relationship 
\begin{equation} 
    S_h(f) = \frac{3H^2_0}{4\pi ^2 f^3}\Omega_{\mathrm{gw}}(f).
\end{equation}

The classical Hamiltonian described the tidal interaction between the SGWB and a hydrogen atom is
\begin{align}
    H_I = &\frac{1}{4}\mu r^2 \times\notag\\ 
    &\sum\limits_{A=+,\times}\int d^2\hat{n}\;\hat{x}^j\hat{x}^ke^A_{jk}(\hat{n}) \int df\; (2\pi f)^2\tilde{h}_A(f,\hat{n})e^{-2\pi fti},
\end{align}
where $r^2 \equiv \delta_{jk}x^jx^k$ and $\hat{x}^i = x^i/r$. From this one can view the atom as a tensor operator, denoted by $D^{jk}$, that maps the GW tensor $\ddot{h}^{\mathrm{TT}}_{jk}$ onto a scalar quantity representing the interaction Hamiltonian: $H_{I} = D^{jk}h^{\mathrm{TT}}_{jk}$, Here $D^{jk} = (1/4)\mu x^jx^k$. The corresponding \emph{detector pattern function} is given by
\begin{equation}\label{patternfunc}
    F^A(\hat{n}) = \hat{x}^j\hat{x}^k e^A_{jk}(\hat{n}),
\end{equation}
which depends on the wave's propagation direction $(\theta,\phi)$. 

The effective interaction Hamiltonian is
\begin{equation}
    H^{\mathrm{eff}}_{I} = \sqrt{\langle H^2_I(t)\rangle} = \frac{1}{4}\mu r^2 F \left[\int_0^\infty df\, (2\pi f)^2S_h(f)\right]^{1/2}
\end{equation}
where
\begin{align}\label{angeff}
    F^2 \equiv \sum\limits_{A=+,\times} \int_{0}^{2\pi} \frac{d\psi}{2\pi} \int \frac{d^2\mathbf{n}}{4\pi} F^A(\hat{n},\psi)F^A(\hat{n},\psi).
\end{align}
By straightforward calculation, described in Appendix \ref{angfac}, we find that $F^2 = 8\pi/15$.  

Using this result, the energy shifts induced by the SGWB can be computed via perturbation theory, as discussed previously.

\section{Conclusion and outlook}
This paper examines how GWs affect hydrogen radio-frequency spectral lines, particularly through hyperfine splitting and broadening in RRLs. Using both quantum-mechanical and semi-classical methods, we show that GWs cause potentially detectable energy shifts in highly excited states of hydrogen atoms, resulting in spectral broadening that the fractional linewidth scales with $n^7\omega^2_{\mathrm{gw}}h(t)$. This suggests that high-$n$ RRLs could serve as astrophysical probes for ultra-high-frequency GWs.

Our theoretical analysis indicates that GW-induced broadening could dominate over natural broadening in specific astrophysical environments, particularly for inspiraling primordial black hole binaries with planetary masses. While present-day radio telescopes may lack the resolution to observe this effect, next-generation spectroscopic facilities like the SKA should be able to achieve such detectability. Additionally, studying the impact of a stochastic gravitational-wave background on hydrogen spectral lines offers a potential avenue for indirect GW detection.

This work establishes atomic spectroscopy as a promising tool for gravitational-wave astronomy. Future studies could refine these predictions by incorporating interstellar environmental factors, enhancing observational techniques, and investigating potential GW signatures from astrophysical sources.

\appendix
\begin{acknowledgments}
This research has received funding support from Fundamental Fund 2023, Chiang Mai University and from the National Science, Research and Innovation Fund (NSRF) via the Program Management Unit for Human Resources \& Institutional. Development, Research and Innovation [grant number B39G680007]. We also thank Prof. Hoi Lai Yu for his valuable comments and suggestions.
\end{acknowledgments}

\section{The angular matrix elements ($\mathcal{F}^{mm'}_{\ell\ell'}$)}\label{appendixA}
The angular matrix element $\mathcal{F}^{mm'}_{\ell\ell'}$ defined in Eq. \eqref{angmatrix}. These matrix elements can be evaluated using the integral
    \begin{align}
     \langle \ell m| Y^{\pm 2}_2| \ell' m'\rangle = \int Y^{*m}_{\ell} (\theta,\phi) Y^{\pm 2}_{2}(\theta,\phi)Y^{m'}_{\ell'}(\theta,\phi) d\Omega \notag\\
     = (-1)^{-m}\sqrt{\frac{5}{4\pi}(2\ell + 1)(2\ell ' +1)} \left(\begin{matrix}
\ell&2&\ell'\\
0&0&0
\end{matrix}\right)
\left(\begin{matrix}
\ell &2&\ell'\\
m&\pm 2&m'
\end{matrix}\right)
\end{align}
where the $2\times 3$ matrices are the Wigner $3j$-symbols. The non-zero elements of $\mathcal{F}^{m,m'}_{\ell\ell'}$ are given by
\begin{widetext}
\begin{align}
    \mathcal{F}^{m,m+2}_{\ell\ell} &= - \frac{1}{(2\ell -1)(2\ell + 3)} \sqrt{\frac{(\ell - m)!(\ell + m + 2)!}{(\ell + m)!(\ell - m -2)!}}\\
    \mathcal{F}^{m,m-2}_{\ell\ell} &= - \frac{1}{(2\ell -1)(2\ell + 3)} \sqrt{\frac{(\ell + m)!(\ell - m + 2)!}{(\ell - m)!(\ell + m -2)!}}\\
    \mathcal{F}^{m,m+2}_{\ell,\ell +2} &= \frac{1}{2(2\ell + 1)}\sqrt{\frac{(\ell + m + 4)!}{(\ell + m)!(2\ell + 1)(2\ell + 5)}}\\
    \mathcal{F}^{m,m - 2}_{\ell,\ell +2} &= \frac{1}{2(2\ell + 3)}\sqrt{\frac{(\ell - m + 4)!}{(\ell - m)!(2\ell + 1)(2\ell + 5)}}\\
    \mathcal{F}^{m,m+2}_{\ell,\ell -2} &= \frac{1}{2(2\ell + 1)}\sqrt{\frac{(\ell + m + 4)!}{(\ell + m)!(2\ell + 1)(2\ell + 5)}}\\
    \mathcal{F}^{m,m - 2}_{\ell,\ell +2} &= \frac{1}{2(2\ell + 3)}\sqrt{\frac{(\ell - m)!}{(\ell - m - 4)!(2\ell -3)(2\ell + 1)}}
\end{align}
\end{widetext}

\section{Evaluation of radial matrix elements of the form $\langle n\ell|r^2|n\ell'\rangle$}\label{appendixB}
To calculate the radial quadrupolar matrix element for the hydrogen atom, we consider the expression
\begin{equation}\label{radialelem}
    \langle n\ell|r^2|n \ell'\rangle = \int_0^\infty r^4R_{n\ell}(r)R_{n\ell '}(r)dr
\end{equation}
where $R_{n\ell}$ represents the radial wavefunction of hydrogen,
\begin{equation}
    R_{n\ell}(r) = N_{n\ell} x^\ell e^{-x/2}L^{2\ell+1}_{n-\ell-1}(x)
\end{equation}
with $x \equiv 2r/na$ and $N_{n\ell}$ being the normalization constant:
 \begin{equation} 
 N_{n\ell} = \sqrt{\left(\frac{2}{na}\right)^3\frac{(n-\ell-1)!}{2n(n + \ell)!}}. 
 \end{equation}
Evaluating the matrix element in Eq. \eqref{radialelem} involves computing an integral of the form
 \begin{equation}\label{integral}
     \int_0^\infty e^{-x}x^{\ell+\ell'} x^4 L^{2\ell+1}_{n-\ell-1}(x)L^{2\ell'+1}_{n-\ell'-1}(x)dx.
\end{equation}
For this purpose, the following recursion relation for the associated Laguerre polynomials is useful (reference: \cite{morse1946}):
\begin{equation} 
x L^p_q = (p + 2q + 1) L^p_q - (q + 1) L^p_{q + 1} - (p + q) L^p_{q-1} 
\end{equation}
along with the orthogonality relation
\begin{equation} 
\int_0^\infty e^{-x} x^k  L^k_m(x) L^k_n(x)dx = \frac{(k + m)!}{m!} \delta_{mn}. 
\end{equation}
Using these relations, one can systematically simplify and evaluate the integral in Eq. \eqref{integral}, ultimately leading to an explicit expression for the desired radial matrix element \eqref{radialelem}. 

The radial matrix elements used to construct the perturbation matrix \eqref{Hqm} are
\begin{align}
    \langle n\ell|r^2|n\ell\rangle &= \frac{5a^2n^2}{2}\left[n^2 + \frac{1 - 3\ell(\ell+1)}{5}\right]\\
    \langle n\ell|r^2|n,\ell+2\rangle &= \frac{5a^2n^2}{2}\sqrt{\frac{(n-\ell-1)!(n + \ell + 2)!}{(n + \ell)!(n - \ell -3)!}}\\
    \langle n\ell|r^2|n,\ell-2\rangle &= \frac{5a^2n^2}{2}\sqrt{\frac{(n+\ell)!(n- \ell +1)!}{(n - \ell -1)!(n + \ell -2)!}}
\end{align}

\section{Computing the angular efficiency factor}\label{angfac}
We provide a brief guide to computing the detector pattern function $F^A(\hat{n})$ and the angular efficiency factor $F^2$ for the hydrogen atom as a probe of SGWB described by Eq. \eqref{SGWB}. In spherical polar coordinates, we can write
\begin{align}
    \hat{n} &= \left( {\begin{array}{*{20}{c}}
\sin\theta\sin\phi\\
\sin\theta\cos\phi\\
\cos\theta
\end{array}} \right),\; \hat{x} = \left( {\begin{array}{*{20}{c}}
\sin\theta'\sin\phi'\\
\sin\theta'\cos\phi'\\
\cos\theta'
\end{array}} \right),\\
\mathbf{u} &= \left( {\begin{array}{*{20}{c}}
-\cos\theta\sin\phi\cos\psi - \cos\phi\sin\psi\\
-\cos\theta\cos\phi\cos\psi + \sin\phi\sin\psi\\
\sin\theta\cos\phi
\end{array}} \right),\\
\mathbf{\hat{v}} &=
\left( {\begin{array}{*{20}{c}}
\cos\theta\sin\phi\sin\psi - \cos\phi\cos\psi\\
\cos\theta\cos\phi\sin\psi + \sin\theta\cos\psi\\
-\sin\theta\sin\psi
\end{array}} \right).
\end{align}
Using Eq. \eqref{patternfunc} together with the expression for the polarization tensor $e^A_{jk}$, given in Eq. \eqref{polarizations}, we obtain the pattern functions
\begin{widetext}
    \begin{align}
 F^+(\theta,\phi,\psi) &=  \sin 2 \psi \left[\sin2 \theta' \sin\theta \sin(\phi' - \phi) - \sin^2\theta' \cos\theta \sin 2 (\phi'-\phi) \right] \notag\\ 
 &-\frac{1}{4} \cos 2\psi \left[\sin^2\theta' (\cos 2\theta + 3) \cos 2 (\phi' - \phi)
  -2 \sin 2 \theta' \sin 2\theta \cos (\phi' - \phi) + (3 \cos 2 \theta' + 1) \sin ^2\theta\right],\\
  F^{\times}(\theta,\phi,\psi) &= - \sin 2 \psi \left[ \sin^2\theta' (\cos 2 \theta + 3) \cos [2 (\phi' - \phi) - 
    2 \sin 2\theta' \sin 2 \theta \cos(\phi' - \phi) + (3 \cos 2 \theta' + 1) \sin^2 \theta )\right]\notag\\
    &+ \frac{1}{4} \left[\cos 2 \psi (4 \sin^2\theta' \cos\theta \sin 2(\phi' - \phi) - 4 \sin 2\theta' \sin\theta\sin(\phi' - \phi)\right].
\end{align}
\end{widetext}
Plugging the above results into Eq. \eqref{angeff}, yields the corresponding angular efficiency factor
\begin{equation}
    F^2 = \int_{0}^{2\pi}\frac{d\psi}{2\pi} \int \frac{d^2\hat{n}}{4\pi} [(F^+)^2 + (F^\times)^2] = \frac{8\pi}{15}.
\end{equation}

\nocite{*}

\bibliography{apssamp}% Produces the bibliography via BibTeX.

\end{document}